\documentclass[12pt]{article}
\usepackage[eqsecnum]{pre2e}
\usepackage{gcom,amsfonts,amssymb}
\title[Center-of-mass variables\hfill R.Gaida, V.Tretyak, Yu.Yaremko]
{Quasi-relativistic 
center-of-mass variables in applications}
\author[R.Gaida, V.Tretyak, Yu.Yaremko]%
{\fbox{R.Gaida}, V.Tretyak, Yu.Yaremko}
\address{Institute for Condensed Matter Physics of the
\newline National Academy of Sciences of Ukraine
\newline 1 Svientsitskii St., UA--290011 Lviv, Ukraine}
\date{ }
\renewcommand{\d}{{\rm d}}

\begin{document}
\setcounter{page}{1}
\maketitle
\begin{abstract}
Collective center-of-mass variables are introduced in the
Lagrangian formalism  of  the  relativistic  classical  mechanics  of
directly interacting particles. It is shown that the  transition
to the  Hamiltonian
formalism leads to  the  Bakamjian-Thomas  model. The quantum-mechanical
system consisting of two spinless particles is investigated.
Quasi-relativistic corrections to the discrete energy spectrum
are calculated for some Coulomb-like interactions having
field theoretical analogues.
\pacs 03.20.+i 
\end{abstract}

\section{Introduction}

There are two types of classical relativistic  direct interaction theories
(RDIT), namely, manifestly Poincar\'e-invariant  four-dimensional
formulations and three-di\-men\-si\-o\-nal approaches \cite{GN}. Applying 
each of them to the description of real physical $N$-particle systems we come
across the problem of space-time interpretation of some formal constructions.
In four-dimensional approaches difficulties are caused by the presence of
redundant variables specific to many-time formalisms and in
three-dimensional ones they are caused by the so-called no-interaction
theorem. In the relativistic Hamiltonian mechanics, this theorem forbids
identification (in the interaction zone) of canonical variables $q_a{}^i$
with physical (covariant) position coordinates $x_a{}^i$ ($a=\overline{1,N}$;
$i=1,2,3$) which transform as a space part of the events in the Minkowski
space-time \cite{CJS}. In the Lagrangian formulation
\cite{GN,GKT801,GKT802,GKT_Fl} the no-interaction theorem  asserts
that the interaction Lagrangian depends inevitably on  all  higher
time derivatives $\stackrel{s}{x_a}{}^{\hspace{-3pt}i}=
\d^sx_a{}^i/\d t^s$,
$s=1,2\ldots $,  of the physical  particle coordinates. This theorem
is essentially related to the use of individual particle variables
(PV) in both the Hamiltonian and Lagrangian approaches.

In papers \cite{GN,GKT801,GKT802,GKT_Fl,GT_pl} the following approach
to the construction of RDIT was proposed. The starting point is the
3-dimensional Poincar\'e-invariant Lagrangian formalism \cite{GN,GKT801}
with using the physical coordinates $x_a{}^i$ and with allowing the
connection to field theories via the Fokker-type formalism \cite{GT_pl}.
These aspects of the Lagrangian description along with the possibility to use
the N\"other theorem for obtaining the conservation laws \cite{GKT802} are
the  main advantages of this approach from the point of view of the physical
interpretation of the theory. The next step is a transition to the
canonical Hamiltonian formalism admitting the quantization of the
theory \cite{GKT_Fl,GT_pr}.

Another aspect of RDIT is the problem of how to separate the
motion of the system as a whole from the internal  motion  of  its
constituents. This problem can be solved in a manner similar to that of
the non-relativistic mechanics case by introducing center-of-mass
variables (CMV). Relativistic generalization of the non-relativistic
notion of the center-of-mass is ambiguous. The most frequently used are,
firstly, the center-of-inertia (covariant center-of-mass) whose coordinates
$R^i$ are transformed by the Poincar\'e group as 3-coordinates of the
Minkowski-space points, and, secondly, the center-of-spin (or canonical
center-of-mass) with the canonical coordinates $Q^i$ \cite{Pryce,Flem}.
External variables $Q^i$ and $P_i$ describing the motion of a system
as a whole, along with the corresponding set $(q_b{}^i, p_{bi})$,
$b=\overline{1,N-1}$ of $6(N-1)$ internal variables are often used in
the relativistic Hamiltonian mechanics \cite{Bak_Th}-\cite{Dv_Kl}.

In the relativistic Lagrangian formalism it is natural to
choose the coordinates of the center-of-inertia ${\bf R}$ as external
position variables
and to formulate the problem of constructing the relativistic Lagrangian
mechanics of an $N$-particle  system in terms of CMV.
This set also contains $3(N-1)$ internal variables $\rho_b{}^i$
\cite{GTY_TM}-\cite{GTY_UPJ}. The internal variables $\rho_b{}^i$ may be
introduced by postulating their transformation properties with respect to
the Poincar\'e group. The corresponding relations which are written down
in section 2 constitute the prerequisites which enable us to find a
connection  between CMV and PV and, accordingly, between the relativistic
Lagrangians in terms of these two classes of variables. An important feature
of CMV  is the existence of the Poincar\'e-invariant interaction Lagrangians
which depend only on a finite number of time derivatives (section 3). The
transition to the Hamiltonian formalism (section 4) leads to the
well-known Bakamjian-Thomas model supplemented by the relations between the
canonical variables and the physical particle  positions.  In  sections 5
and 6 the first-order quasi-relativistic approximation (up to $c^{-2}$
terms) is considered from both the classical and the quantum mechanical
viewpoints. Relativistic corrections to the energy spectrum of a
two-particle system with Coulomb-like interactions are  calculated
as an illustration.
\section{Realization of the Poincar\'e  group ${\cal P}(1,3)$
in terms of the Lagrangian CMV}

Since the  external variables $R^i$ describe the motion of an
$N$-particle system as a whole, we require their transformation
properties with respect to the Poincar\'e group to be the same as those of
the physical coordinates $x_a{}^i$ of a single point particle.
In  the  instant  form  of dynamics \cite{GKT801,GN} they are
\cite{GTY_TM,GTY_UPJ}:
\begin{eqnarray} \label{t_R}
{\cal H}R_i = \dot R_i\,,&& {\cal P}_iR_j = \delta_{ij}\,,\quad
{\cal J}_iR_j = - \varepsilon_{ijk}R_k\,,\\
{\cal K}_iR_j&=&-t\delta_{ij} + c^{-2}R_i\dot R_j\,.
\end{eqnarray}
Here ${\cal H}:=D-\partial /\partial t$, ${\cal P}_i$, ${\cal J}_i$, and
${\cal K}_i$ are generators of time and space translations, space rotations,
and boosts, respectively, and $D$ is the total time derivation operator.
As usual, the dot denotes the time derivation (e.g., $\dot R_i=DR_i$
are the components of the external velocity ${\bf\dot R}$).
We postulate further that similar to the case of non-relativistic mechanics
the internal CMV $\rho_b{}^i$ are the components of the translationally
invariant 3-vectors ${\mbox{\boldmath$\rho$}}_b$
$(b=\overline{1,N-1})$ which do not depend explicitly on time:
\begin{equation} \label{t_r}
{\cal H}\rho_{bi} = \dot\rho_{bi}\,,\quad {\cal P}_i\rho_{bj} = 0\,,\quad
{\cal J}_i\rho_{bj} = - \varepsilon_{ijk}\rho_{bk}\,.
\end{equation}
The transformation of these variables with respect to boosts is
determined by the Poincar\'e group structure. It can be shown that
the commutation relations of the Poincar\'e algebra ${\cal AP}(1,3)$, along
with equations (\ref{t_R})--(\ref{t_r}), lead to the following conditions
\cite{GTY_TM,GTY_UPJ}:
\begin{equation} \label{tr}
{\cal K}_i\rho_{bj} = c^{-2}\left(R_i\dot\rho{_bj} + \dot R_i\rho_{bj} -
\delta_{ij}({\bf\dot R{\mbox{\boldmath$\rho$}}_b})\right)\,.
\end{equation}
The expressions for generators
${\cal X}_\alpha = \{{\cal H},{\cal P}_i ,{\cal J}_i,{\cal K}_i\}$
of the Poincar\'e group in terms of CMV may be obtained from
equations (\ref{t_R})--(\ref{tr}) immediately \cite{GTY_TM}.

\section{Poincar\'e-invariance of the Lagrangian description}

The evolution of a particle system is described by the solutions of
the Euler-Lagrange-Ostrogradsky equations
\begin{equation} \label{EL}
\sum_{s=0}^{\infty}(-D)^s\frac{\partial L}{\partial
\stackrel{s}{z}_\nu} = 0\,,\quad \{z_\nu\} = \{R_i, \rho_{bj}\},
\end{equation}
following from the variational principle $\delta\int L\d t =  0$.
The conditions of Poincar\'e invariance of equations (\ref{EL}) may be
expressed by the system of first-order differential equations for the
Lagrange function:
\begin{equation} \label{Pi}
{\cal X}_\alpha L = D\Omega_\alpha \,,
\end{equation}
where the auxiliary functions $\Omega_\alpha$ have to satisfy the
compatibility conditions:
\begin{equation} \label{cc}
{\cal X}_\alpha\Omega_\beta - {\cal X}_\beta\Omega_\alpha =
c_{\alpha\beta}{}^\gamma\Omega_\gamma\,,
\end{equation}
with $c_{\alpha\beta}{}^\gamma$ being the structure constant tensor of the
Poincar\'e group ${\cal P}(1,3)$. For this group the relations (\ref{cc})
admit the following set of functions $\Omega_\alpha$:
\begin{equation} 
\Omega_{\cal H} = L\,,\quad \Omega_{{\cal P}_i} = 0\,,\quad
\Omega_{{\cal J}_i} = 0\,,\quad \Omega_{{\cal K}_i} = c^{-2}R_iL\,.
\end{equation}
In which case the conditions (\ref{Pi}) of Poincar\'e invariance take on
the form:
\begin{eqnarray} \label{Cc}
{\cal P}_iL = 0\,,&& {\cal J}_iL = 0\,,\quad
({\cal H} - D)L\equiv - \frac{\partial L}{\partial t} = 0\,,\nonumber\\
&&({\cal K}_i - c^{-2}R_iD)L - c^{-2}\dot R_iL = 0\,.
\end{eqnarray}

The structure of equations (\ref{Cc}) is quite different
from the structure of the Poincar\'e invariance conditions
in terms of  PV $x_a{}^i$ \cite{GKT801,GKT802}. Whereas the interaction
Lagrangian as a function of the particle variables is defined
on $J^\infty\pi$, equations (\ref{Cc}) admit Lagrangians that
depend on CMV and their time derivatives of finite  orders.
The simplest structure of $L$ is given by
\begin{equation} \label{L}
L({\bf\dot R},{\bf\ddot R},
{\mbox{\boldmath$\rho$}}_b , {\mbox{\boldmath$\dot\rho$}}_b) =
\Gamma^{-1}F({\mbox{\boldmath$\alpha$}}_b,
{\mbox{\boldmath$\beta$}}_b,
{\mbox{\boldmath$\gamma$}})\,, b=\overline{1,N-1};
\end{equation}
here  $\Gamma^{-1}=\sqrt{1-c^{-2}{\bf\dot R}^2}$ and $F$ is an
arbitrary function of scalar products of the vectors
\begin{eqnarray}
{\mbox{\boldmath$\alpha$}}_b&=&{\mbox{\boldmath$\rho$}}_b +
c^{-1}\Gamma[{\bf\dot R} {\mbox{\boldmath$\rho$}}_b]\,,\qquad
{\mbox{\boldmath$\beta$}}_b = \Gamma{\mbox{\boldmath$\dot\alpha$}}_b -
c^{-1}\Gamma^2[{\bf\ddot R} {\mbox{\boldmath$\alpha$}}_b]\,,\\
{\mbox{\boldmath$\gamma$}}&=&\Gamma{\bf\ddot R} + c^{-1}\Gamma^2[{\bf\dot
R}{\bf\ddot R}]+c^{-2}\Gamma({\bf\dot R}{\bf\ddot R}) {\bf\dot R}\,.
\end{eqnarray}
Contrary to ${\mbox{\boldmath$\alpha$}}_b$  and
${\mbox{\boldmath$\beta$}}_b$ , the ${\mbox{\boldmath$\gamma$}}$
dependence of $L$ is not necessary. The Lagrangian (\ref{L}) depending
on  a  finite  number  of  CMV  can  be made
equivalent to the Lagrangian defined on an infinite-dimensional
set of PV $\stackrel{s}{x}_a{}^i$, only if the CMV are functions
of all these PV. The set of equations (\ref{t_R})--(\ref{tr}) with
generators ${\cal H}$, ${\cal P}_i$, ${\cal J}_i$, and ${\cal K}_i$,
determined in terms of PV \cite{GKT801,GKT802}, may be regarded as
the basis for determining these functions \cite{GTY_UPJ}.

Using the N\"other theorem for Lagrangians with higher derivatives
\cite{GN}-\cite{GKT802} we found ten integrals of motion
$(E,{\bf P},{\bf J},{\bf K})$ corresponding to the Poincar\'e-invariance
of the theory, based on Lagrangian (\ref{L}). From
their analysis we obtain \cite{GTY_TM,GTY_UPJ} the following equation
of  external motion
\begin{equation} \label{Em}
M{\bf R} + c^{-2}[{\bf\dot R},{\bf s}] = {\bf K}_0
\end{equation}
in the rest frame of reference ($E=E_0=Mc^2$, ${\bf P}=0$, ${\bf J}={\bf s}$,
${\bf K}={\bf K}_0)$).
It should be noted that the equation
\begin{equation} \label{eme}
M{\bf\ddot R} + c^{-2}[{\bf\stackrel{...}{\bf R}}, {\bf s}] = 0\,,
\end{equation}
being the consequence of (\ref{Em}), was obtained in \cite{WR}
for  classical particles with a spin; it is a three-dimensional form of
the well-known Mathisson equation. Spin {\bf s} is equal to the
internal angular momentum of the system.

Solving (\ref{Em}) in a coordinate system with the origin at point
${\bf R}_0={\bf K}_0/M$, where ${\bf K}_0$ is the value of integral
${\bf K}$ in the rest frame and the O$z$ axis is directed along spin ${\bf s}$,
we see that point ${\bf R}(t)$ describes a circular motion in a plane
orthogonal to the spin:
\begin{equation} \label{circ}
     R_x(t) = A\sin(\omega_0t - \alpha )\,,\quad
R_y(t) = A\cos(\omega_0t - \alpha )\,,\quad  R_z  = 0\,.
\end{equation}
Here, frequency $\omega_0=Mc^2/s$; $A$ and  $\alpha$ are constants
of the integration. Taking into account the usual non-relativistic  limit
procedure, in accordance with which the center of mass is at rest
as measured in the rest frame, we must set $A = 0$. Indeed, as
$c^{-2}\to 0$ ($\omega_0\to\infty $), the limits of the trigonometric
functions in equation (\ref{circ}) do not exist.
This means that in the set of solutions of equations (\ref{EL}) for the
Lagrangian (\ref{L}) we keep only those which satisfy  the
standard free-particle equation ${\bf\ddot R}=0$.

\section{The Hamiltonian description}

There are several ways to  obtain
the Hamiltonian formulation corresponding to the Lagrangians  with
higher  derivatives  and  with  supplementary  conditions,  like
${\bf\ddot R}=0$ \cite{GT_pr,DS,JLM}. Here we use an approach
close to that of \cite{DS}. Let us perform a change of the variables
$({\bf R}, {\mbox{\boldmath$\rho$}}_b) \mapsto
({\bf Q},{\bf q}_b)$, where
\begin{eqnarray} \label{QR}
{\bf Q}&=&{\bf R} + \frac{[{\bf P, s}]}{M(E+Mc^2)}\,,\\
{\bf q}_b&=&\Gamma{\mbox{\boldmath$\rho$}}_b -
\frac{c^{-2}\Gamma^2}{\Gamma +
1}{\bf\dot R}({\bf\dot R}{\mbox{\boldmath$\rho$}}_b)\,.
\end{eqnarray}
Relation (\ref{QR}) is a definition of  the  canonical  center-of-mass
\cite{Pryce}. In terms of new variables, the Lagrangian (\ref{L}) has the
form:
\begin{equation} \label{L_td}
L({\bf\dot R},{\bf\ddot R},
{\mbox{\boldmath$\rho$}}_b , {\mbox{\boldmath$\dot\rho$}}_b) =
L({\bf\dot Q}, 0, {\bf q}_b, {\bf\dot q}_b) + L' \equiv
\tilde L({\bf\dot Q}, {\bf q}_b, {\bf\dot q}_b) + L'\,,
\end{equation}
where  $L'$  contains only the  terms which do  not  contribute to
the equations of motion if the external equation of motion
${\bf\ddot R}=0$ is satisfied. Therefore, the term $L'$ may be omitted in
equation (\ref{L_td})  and,  thus,  the   transformation   to   the
Hamiltonian description may be performed in the usual manner on the  basis
of Lagrangian  $\tilde L$. Using the Legendre transformation we construct
Hamiltonian  $H$  corresponding to Lagrangian $\tilde L$:
\begin{equation} \label{Hbk}
H = \sqrt{c^2{\bf P}^2 + M^2({\bf q}_b, {\mbox{\boldmath$\pi$}}_b)c^4}\,,
\end{equation}
where
\begin{equation} 
{\bf P} = \frac{\partial\tilde L}{\partial{\bf\dot Q}}\,,\quad
{\mbox{\boldmath$\pi$}}_b =
\frac{\partial\tilde L}{\partial{\bf\dot q}_b}\,,\quad
Mc^2 = \left.\sum_{b=1}^{N-1}{\mbox{\boldmath$\beta$}}_b\frac{\partial
F}{\partial {\mbox{\boldmath$\beta$}}_b} - F\right|_{{\bf\ddot R}=0}\,.
\end{equation}
We note that the external momentum variable ${\bf R}$ is a motion
integral (the total momentum of the system).

Expression (\ref{Hbk}) for $H$ reproduces the particle-system
Hamiltonian of the Ba\-kam\-jian-Thomas model \cite{Bak_Th}. The advantage of
our way of constructing this model is the possibility  to  express
the internal Hamiltonian
$M({\bf q}_b, {\mbox{\boldmath$\pi$}}_b)$
in  terms  of  the  interaction
Lagrangian which, in turn, may be related  to  some  field  models
\cite{GT_pl}. On the other hand, in our approach the canonical variables
${\bf Q}, {\bf P}, {\bf q}_b$, and ${\mbox{\boldmath$\pi$}}_b$
have a definite relation to the Lagrangian CMV
${\bf R}, {\mbox{\boldmath$\rho$}}_b$ and their time derivatives
${\bf\dot R}, {\mbox{\boldmath$\dot\rho$}}_b$
and, therefore, to the individual particle variables ${\bf x}_a$, at
least in the approximations in the parameter $c^{-2}$.

\section{Quasi-relativistic approximation}

Let us consider the main steps of our approach in the first-order
approximation in $c^{-2}$ (i.e. quasi-relativistic  or post-Newtonian
approximation).  We use the following sufficiently  general
form  of a quasi-relativistic two-particle Lagrangian with
static non-relativistic potential $U(r)$ \cite{WH,GN,GKT802}:
\begin{eqnarray} \label{Lqr}
L&=&\sum_{a=1}^{2}\left(\frac{m_av_a^2}{2} +
\frac{m_av_a^4}{8c^2}\right) - U(r) -
\frac{1}{2c^2}\left((-{\bf v}_1{\bf v}_2 + 2Av^2)U(r) +\right.\nonumber\\
&+&\left.\left(({\bf rv}_1)({\bf rv}_2) + 2B({\bf rv})^2\right)
\frac{1}{r}\frac{\d U}{\d r} + 2u_1 (r)\right)\,,
\end{eqnarray}
where ${\bf r}={\bf x}_1 -{\bf x}_2, {\bf v}={\bf v}_1 -{\bf v}_2$,
$A$ and $B$ are arbitrary constants  and $u_1(r)$
is an arbitrary function. For interactions which  correspond  to  the
linear field theories, we have  $u_1(r)=0$  and  constants  $A$  and  $B$
depend on the tensor structure of interaction \cite{GN,WH}.  For
example, we have $A=B=0$ for a vector interaction, and  $A=-1,  B=0$
for a scalar interaction.

Further we consider a specific case of interactions which correspond
to the non-relativistic long-range potential
\begin{equation} \label{U(r)}
U(r) = \frac{\alpha}{r}\,,
\end{equation}
where $\alpha$ is a constant of the interaction. We suppose that the
Galileo-invariant function $u_1(r)$ has the form:
\begin{equation} \label{u_1}
u_1(r) = \frac{\alpha_1}{r^2}\,,
\end{equation}
where $\alpha_1$ is the constant. Let us note five significant cases
covered by the interaction  described in a quasi-relativistic
approximation by the Lagrangian of type (\ref{Lqr}) where functions
(\ref{U(r)}) and (\ref{u_1}) enter:
\begin{itemize}
\item
$\alpha =e_1e_2, A=B=0$, and $\alpha_1=0$: the Darwin Lagrangian which
describes an electromagnetic interaction of two point charges;
\item
$\alpha =-Gm_1m_2, 2A=3, B=0$, and $2\alpha_1=G^2m_1m_2m (m=m_1+m_2)$ :
the Einstein-Infeld-Hoffmann Lagrangian which
describes a gravitational interaction of two point masses;
\item
$\alpha =e_1e_2-Gm_1m_2, 2A=-3Gm_1m_2/\alpha, B=0$, and
$2\alpha_1=G(Gm_1m_2m+e_1^2m_2+e_2^2m_1-2e_1e_2m)$:
a gravitational interaction of charged bodies (the Ba\-\.za\'n\-ski
Lagrangian \cite{Bz});
\item
$\alpha =-Gm_1m_2, 2A=1+2\gamma, B=0$, and
$2\alpha_1=(2\beta-1)G^2m_1m_2m$:
a gravitational interaction in the PPN-formalism \cite{Kl} which
classifies suitable theories of gravitation (i.e.\ those which do not
contradict the four classical experiments, satisfy the conditions of the
Lorentz-invariance, and admit $N$-particle Lagrangian functions)
in the first-order approximation in $c^{-2}$;
\item
$\alpha =-Gm_1m_2, A=-2B=1$, and $\alpha_1=0$: the Lagrangian of
the quasi-relativistic approximation of the Whitehead-Shield theory of
gravitation \cite{WM} where the massless tensor field of the second rank
as an intermediate field is assumed.
\end{itemize}
The correspondence between CMV and PV for a  two-particle  system
in the first-order quasi-relativistic approximation has the following
form \cite{GTY_UPJ} :
\begin{eqnarray} \label{xRr}
{\bf x}_a&=&{\bf R}+(-1)^b\frac{m_b}{m}{\mbox{\boldmath$\rho$}}
+\frac{1}{c^2}\left\{(-1)^b\frac{m_b}{m}\left(\frac{1}{2}{\bf\dot R}^2
{\mbox{\boldmath$\rho$}} -
({\bf\dot R}{\mbox{\boldmath$\rho$}}){\bf\dot R}\right)-\right.\nonumber\\
&-&\left.\frac{m_b^2}{m^2}({\bf\dot R}{\mbox{\boldmath$\rho$}})
{\mbox{\boldmath$\dot\rho$}} + \frac{m_1-m_2}{2m^2}\left(\mu{\mbox{\boldmath$\dot\rho$}}^2 +
U(\rho)\right){\mbox{\boldmath$\rho$}}\right\}\,,
\end{eqnarray}
where $a,b=1,2, b\ne a, m=m_1+m_2, \mu=m_1m_2/m$. Using these relations
in the Lagrangian (\ref{Lqr}) leads to the following two-particle
quasi-relativistic Lagrangian in terms of CMV:
\begin{eqnarray} \label{LRr}
L&=&\frac{m{\bf\dot R}^2}{2}+\frac{\mu{\mbox{\boldmath$\dot\rho$}}^2}{2}
-U(\rho)+\frac{m{\bf\dot R}^4}{8c^2}+
\frac{\mu{\mbox{\boldmath$\dot\rho$}}^4}{8c^2}\left(1-\frac{3\mu}{m}\right)
+ \frac{3\mu}{4c^2}{\bf\dot R}^2{\mbox{\boldmath$\dot\rho$}}^2 -\nonumber\\
&-&\frac{\mu}{2c^2}({\bf\dot R}{\mbox{\boldmath$\dot\rho$}})^2 +
\frac{\mu}{c^2}\left({\bf\dot R}\bigl[{\mbox{\boldmath$\rho$}}[{\bf\ddot
R}{\mbox{\boldmath$\dot\rho$}}]\bigr]\right)+\frac{1}{2c^2}\left\{
\left({\bf\dot R}^2-\left(\frac{\mu}{m} +
2A\right){\mbox{\boldmath$\dot\rho$}}^2\right)U(\rho) +\right.\nonumber\\
&+&\left.\left(\left(\frac{\mu}{m}-2B\right)({\mbox{\boldmath$\rho$}}{\mbox{\boldmath$\dot\rho$}})^2
- [{\bf\dot R}{\mbox{\boldmath$\rho$}}]^2\right)
\frac{1}{\rho}\frac{\d U}{\d\rho}\right\} - \frac{1}{c^2}u_1(\rho)\,.
\end{eqnarray}
This expression agrees with the corresponding approximation of
the La\-gran\-gian (\ref{L}).

In the quasi-relativistic approximation the canonical external variable
(\ref{QR}) becomes
\begin{eqnarray} \label{aQR}
{\bf Q}&=&{\bf R} + \left.\frac{[{\bf P,
s}]}{M(E+Mc^2)}\right|_{{\bf\ddot R}=0}\equiv\nonumber\\
&\equiv&{\bf R} +
\frac{\mu}{2mc^2}\left[{\bf\dot R}[{\mbox{\boldmath$\rho$}}{\mbox{\boldmath$\dot\rho$}}]\right]\,.
\end{eqnarray}
The total momentum
\begin{equation}
{\bf P} = \left.{\bf P}\right|_{{\bf\ddot R}=0}\equiv\frac{E}{c^2}{\bf\dot R}
\end{equation}
is canonically conjugated to {\bf Q}. Quasi-relativistic inner canonical
variables are  defined  by   the relations
\begin{eqnarray} 
{\bf q}&=&{\mbox{\boldmath$\rho$}} +
\frac{\left[{\bf\dot R}[{\mbox{\boldmath$\rho$}}{\mbox{\boldmath$\dot\rho$}}]\right]}{2c^2}
+ \frac{1}{\mu c^2}\frac{\partial\lambda}{\partial
{\mbox{\boldmath$\dot\rho$}}}\,,\\
{\mbox{\boldmath$\pi$}}&=&\left(1-\frac{{\bf\dot R}^2}{2c^2}\right)
\frac{\partial L}{\partial {\mbox{\boldmath$\dot\rho$}}} -
\frac{1}{c^2}\frac{\partial\lambda}{\partial{\mbox{\boldmath$\rho$}}}\,,
\end{eqnarray}
where $\lambda({\mbox{\boldmath$\rho$}},{\mbox{\boldmath$\dot\rho$}})$
is  an arbitrary  Galileo-invariant  function.  Choosing
this function in the form
\begin{equation}
\lambda({\mbox{\boldmath$\rho$}},{\mbox{\boldmath$\dot\rho$}}) =
({\mbox{\boldmath$\rho$}}{\mbox{\boldmath$\dot\rho$}})
\left(\frac{\mu{\mbox{\boldmath$\dot\rho$}}^2}{8}\left(1-\frac{3\mu}{m}\right)
+ \frac{1}{4}\left(1-\frac{\mu}{m}-4B\right)U(\rho)\right)\,,
\end{equation}
we obtain the Hamiltonian function depending on the square of the inner
momentum ${\mbox{\boldmath$\pi$}}$ and therefore being convenient for
quantization:
\begin{equation} \label{H}
H = \frac{{\bf P}^2}{2m} - \frac{{\bf P}^2}{2m^2c^2}
\left(\frac{{\mbox{\boldmath$\pi$}}^2}{2\mu} + U(q)\right)
- \frac{{\bf P}^4}{8m^3c^2} + \frac{{\mbox{\boldmath$\pi$}}^2}{2\mu(q)}
+ v(q)\,.
\end{equation}
Here
\begin{equation} \label{a(q)}
\frac{1}{\mu(q)} = \frac{1}{\mu} + \frac{1}{4\mu^2c^2}\left\{
2\left(1+\frac{\mu}{m}+4A-4B\right)U(q) -
\left(1-\frac{3\mu}{m}\right)q\frac{dU}{dq}
\right\}
\end{equation}
and
\begin{equation} \label{v(q)}
v(q) = U(q) + \frac{1}{c^2}\left\{u_1(q) -
\frac{1}{4\mu}\left(1-\frac{\mu}{m}-4B\right)U(q)q\frac{dU}{dq}
\right\}\,.
\end{equation}

\section{Quasi-relativistic Schr\"odinger equation}
\setcounter{equation}{0}
Starting with the Hamiltonian function (\ref{H}) we write down the stationary
Schr\"odinger equation:
\begin{equation} \label{SE}
(\hat H - E)\langle {\bf Q},{\bf q}\mid E, {\bf P}\rangle = 0\,.
\end{equation}
The total wave function $\langle{\bf Q},\!{\bf q}|E,\!{\bf P}\rangle$
of the
system is equal to the product $\Psi({\bf Q})\psi({\bf q})$ of the external
wave function $\Psi({\bf Q})$ and the inner one $\psi({\bf q})$. We
consider the states with sharply defined total momentum ${\bf P}$,
so the external wave function is then a plane wave,
$\Psi({\bf Q})=\exp(i{\bf kQ})$, where the wave vector
${\bf k}={\bf P}/\hbar$.
Due to the separation of the external and internal variables in
equation (\ref{SE}), we obtain the following wave equation:
\begin{equation} \label{se}
(\hat h - {\cal E})\psi({\bf q}) = 0\,,
\end{equation}
where
\begin{equation} 
{\cal E} = E - \frac{P^2}{2m} + \frac{EP^2}{2m^2c^2} -
\frac{P^4}{8m^3c^2}\,.
\end{equation}
Here $E$ and ${\bf P}$ are eigenvalues of the non-relativistic
Hamiltonian and total momentum operators, respectively.

The internal Hamiltonian operator is defined by the formula
\begin{equation} 
\hat h = \frac{1}{2}\hat X + v(q)\,,
\end{equation}
where the first term is some Hermitian operator which corresponds to
the classical function $\pi^2/\mu(q)\equiv a(q)\pi^2$. We use the following
sufficiently general quantization rule:
\begin{equation} \label{qr}
\hat X =a(q)\hat\pi^2 + \delta^{ij}[\hat\pi_i , a(q)]\hat\pi_j +
\lambda\hat\pi^2(a(q))\,,
\end{equation}
where the inner momentum operator is chosen in the usual form:
\begin{equation} 
\hat\pi_i = -i\hbar\frac{\partial}{\partial q^i}\,;\qquad
\hat\pi^2 = -\hbar^2\Delta\,.
\end{equation}
This rule preserves the rotational invariance of the inner Hamiltonian and
depends on one parameter $\lambda \in \R$. The value $\lambda =1/2$
corresponds to the symmetrization $(\hat\pi^2a(q) + a(q)\hat\pi^2)/2$.

To illustrate our approach we consider the Coulomb-like potential
(see equations (\ref{U(r)}) and (\ref{u_1})). In this case the functions
$a(q):=\mu^{-1}(q)$ and $v(q)$ take the form:
\begin{eqnarray} \label{a}
a(q)&=&\frac{1}{\mu}\left(1 + \frac{1}{\mu c^2}C_1\frac{\alpha}{q}\right)\,,
\quad C_1=\frac{3}{4} - \frac{\mu}{4m} + 2A - 2B\,,\\ \label{v}
v(q)&=&\frac{\alpha}{q}\left(1 + \frac{1}{\mu c^2}C_2\frac{\alpha}{q}\right)\,,
\quad C_2=\frac{1}{4} - \frac{\mu}{4m} - B + \frac{\mu\alpha_1}{\alpha^2}\,.
\end{eqnarray}
Here $\mu=m_1m_2/m$ and the parameters $A$, $B$, $\alpha_1$, and $\alpha$
are defined by the type of interaction (see section 5).

After some calculations we arrive at the following
Hamiltonian operator:
\begin{equation} \label{h}
\hat h = \frac{\hbar^2}{2}\left(-\frac{1}{\mu(q)}\Delta +
\frac{\alpha}{\mu^2c^2}C_1\frac{{\bf q}}{q^3}\frac{\partial}{\partial
{\bf q}} + \frac{4\pi\alpha\lambda}{\mu^2c^2}C_1\delta({\bf q})\right)
+ v(q)\,,
\end{equation}
where $\pi$ denotes a well-known transcendental number instead of the
value of the inner momentum.

We shall study the discrete spectrum of $H$, when $\alpha <0$. The use of
methods of the perturbation theory (e.g.,\ see \cite{Ms}) allows us to
obtain quasi-relativistic corrections to the non-relativistic Bohr
expression
\begin{equation} \label{E0}
{\cal E}_n^{(0)} = - \frac{\mu\alpha^2}{2\hbar^2}\frac{1}{n^2}\,,
\end{equation}
where $n$ is the main quantum number. Let us rewrite the Hamiltonian
operator (\ref{h}) in the form
\begin{equation} \label{h_zb}
\hat h = \frac{\mu}{\mu(q)}\hat h^{(0)} + \frac{C_2-C_1}{\mu c^2}
\frac{\alpha^2}{q^2} + \frac{\hbar^2}{2}
\frac{\alpha}{\mu^2c^2}C_1\left(\frac{{\bf q}}{q^3}\frac{\partial}{\partial
{\bf q}} + 4\pi\lambda\delta({\bf q})\right)\,,
\end{equation}
where
\begin{equation} \label{h_0}
\hat h^{(0)} = - \frac{\hbar^2}{2\mu}\Delta + \frac{\alpha}{q}
\end{equation}
is a non-relativistic energy operator possessing spectrum
(\ref{E0}). According to the stationary perturbation theory \cite{Ms},
the quasi-relativistic term ${\cal E}_{nl}^{(1)}$ is equal to the matrix
element
\begin{equation} \label{me}
{\cal E}_{nl}^{(1)} = \int \d^3q
\psi_{nlm}^{*(0)}({\bf q})V\psi_{nlm}^{(0)}({\bf q})\equiv
\langle V\rangle _0\,,
\end{equation}
where $\psi_{nlm}^{(0)}({\bf q})$ are the eigenfunctions of
operator (\ref{h_0}) and
\begin{equation} \label{Vzb}
V = \frac{1}{\mu c^2}C_1\frac{\alpha}{q}\hat h^{(0)} +
\frac{C_2-C_1}{\mu c^2}\frac{\alpha^2}{q^2} + \frac{\hbar^2}{2}
\frac{\alpha}{\mu^2c^2}C_1\left(\frac{{\bf q}}{q^3}\frac{\partial}{\partial
{\bf q}} + 4\pi\lambda\delta({\bf q})\right)\,,
\end{equation}
is a quasi-relativistic perturbation of the non-relativistic
energy operator \re{h_0}. While calculating (\ref{me}) we take into
consideration that $\psi_{nlm}^{(0)}({\bf q})$ are the
eigenfunctions of $\hat h^{(0)}$. We also use the average values of
the following functions of ${\bf r}={\bf q}$ \cite{Ms}:
\begin{eqnarray} 
&&\langle r^{-1}\rangle _0 = -
\frac{\mu\alpha}{\hbar^2}\frac{1}{n^2}\,,\qquad
\langle r^{-2}\rangle _0 =
\frac{\mu^2\alpha^2}{\hbar^4}\frac{2}{n^3(2l+1)},\nonumber\\
&&\langle \delta({\bf r})\rangle _0
=|\psi_{nlm}^{(0)}(0)|^2=\frac{\mu^3|\alpha
|^3}{\pi\hbar^6n^3}\delta_{0l}\,.
\end{eqnarray}
One can easily check that the
average value is
\begin{equation}
\int \d^3q
\psi_{nlm}^{*(0)}({\bf q})\frac{{\bf q}}{q^3}\frac{\partial}{\partial
{\bf q}}\psi_{nlm}^{(0)}({\bf q})=2\pi|\psi_{nlm}^{(0)}(0)|^2.
\end{equation}

Therefore, the quasi-relativistic correction to the energy spectrum
is given by
\begin{equation} \label{E1}
{\cal E}_{nl}^{(1)} = \frac{\mu\alpha^4}{2\hbar^4c^2}\frac{1}{n^3}
\left(\frac{C_1}{n} - \frac{4(C_1-C_2)}{2l+1}
-(2\lambda -1)\frac{2C_1}{n^3}\delta_{0l}\right)\,.
\end{equation}
We note a dependence of the obtained spectrum on the quantization
parameter $\lambda$.

For the case of an electromagnetic interaction, when the parameters $A=B=0$,
$\alpha_1=0$ (see section~5), spectrum \re{E1} agrees with the results of
the conventional quantum electrodynamics provided that $\lambda =1/2$.
Let us consider a hydrogen-like atom. Then the constant of
interaction $\alpha =-Ze^2$, where $Ze$ is the charge of the nucleus.
Since the mass of the nucleus is much greater than the rest mass of
electron $m_e$, we have $\mu\to m_e$ and, therefore, the
ratio $\mu/m\to 0$.
Here $m$ is the total mass of the atom. Introducing
these parameters into equations (\ref{a}) and
(\ref{v}) we obtain $C_1=3/4$ and $C_2=1/4$. According to equation
(\ref{E1}), the quasi-relativistic correction is equal to
\begin{equation} 
{\cal E}_{nl}^{(1)} = \frac{m_e(Ze^2)^4}{\hbar^4c^2}\frac{1}{n^4}
\left(\frac{3}{8} - \frac{n}{2l+1}\right)\,.
\end{equation}
It is in good agreement with the quasi-relativistic correction
to the hydrogen-like  spectrum which follows from the Dirac equation
(see \cite[p.216]{Ms}).

Now let us consider parapositronium. The mass of an electron
is equal to the mass of a positron and, therefore, $\mu=m_e/2$ and
the ratio $\mu/m=1/4$. Whence the constants $C_1=11/16$ and $C_2=3/16$.
Putting these parameters into equations (\ref{E1}) we re-obtain an
expression for the quasi-relativistic  splitting of the parapositronium spectrum
which  is conventionally found by using the Breit equation \cite{BLP}:
\begin{equation} 
{\cal E}_{nl}^{(1)} = \frac{m_e(e^2)^4}{2\hbar^4c^2}\frac{1}{n^3}
\left(\frac{11}{32n} - \frac{1}{2l+1}\right)\,.
\end{equation}

\section{Conclusions}

The transformation of particle variables into center-of-mass
ones for a class of two-particle quasi-relativistic Lagrangians
with interactions having field-theo\-re\-tical  analogues
has been carried out. As a result, the motion of a system as a whole was
separated from its inner motion. A set of canonical
variables in terms of which the Hamiltonian has a quadratic dependence
on the inner momentum variable is found. The problem is quantized. The
quasi-relativistic splitting of the discrete energy spectrum
is obtained. As expected, the degeneracy in the orbital
quantum number is eliminated.

In the specific case of an electromagnetic interaction between particles,
the quasi-relativistic corrections to the non-relativistic
discrete spectrum are in agreement with the corresponding approximation
of the spectrum of a hydrogen-like atom obtained by using the Dirac
equation (modulus of the spin of the electron which was not taken
into account). For parapositronium (where the summary spin is equal
to zero) the spectrum obtained within the frame of the CMV-formalism
coincides with the one derived by using the Breit equation.


\begin{thebibliography}{99}
\bibitem{GN}
Gaida R.~P. Quasi-relativistic systems of interacting particles // Fiz.\
Elem.\ Chast.\  \& Atom.\ Yadra, 1982, vol.~13, No~2, p.~427--493 (in
Russian; transl.\ in English: Sov.\ J.\ Part. Nucl., 1982, vol.~13, p.~179).
\bibitem{CJS}
Currie D.~G., Jordan J.~F., Sudarshan E.~C.~G. Relativistic invariance
and Hamiltonian theories of interacting particles // Rev.\ Mod.\ Phys.,
1963, vol.~35, No~2, p.~350--375.
\bibitem{GKT801}
Gaida R.~P., Kluchkovsky Yu.~B., and Tretyak V.~I. Lagrangian classical
relativistic mechanics of a system of directly interacting particles.\ I //
Teor.\ Mat.\ Fiz.\ 1980, vol.~44, No~2, p.~194--208 (in
Russian; transl. in English: Theor.\ Math.\ Phys., 1980, vol.~44, p.~687).
\bibitem{GKT802}
Gaida R.~P., Kluchkovsky Yu.~B., and Tretyak V.~I. Lagrangian classical
relativistic mechanics of a system of directly interacting particles.\ II //
Teor.\ Mat.\ Fiz.\ 1980, vol.~45, No~2, p.~180--198 (in
Russian; transl.\ in English: Theor.\ Math.\ Phys., 1980, vol.~45, p.~963).
\bibitem{GKT_Fl}
Gaida R.~P., Kluchkovsky Yu.~B., and Tretyak V.~I. Three-dimensional
Lagrangian approach to the classical relativistic  dynamics of
directly  interacting  particles. -- In:  Constraint's  Theory  and
Relativistic Dynamics. G.~Longhi, L.~Lusanna, editors. World Scientific
Publ., 1987, p.~210--241.
\bibitem{GT_pl}
Gaida R.~P.\ and Tretyak V.~I. Single-time  form  of the  Fokker-type
relativistic dynamics // Acta  Phys.\ Pol.\ B, 1980, vol.~11, No~7,
p.~502--522.
\bibitem{GT_pr}
Gaida R.~P.\ and Tretyak V.~I. Direct interaction Lagrangians and
Hamiltonian description of a particle system in various forms of
relativistic dynamics./ Preprint of the Institute for Theor.\
Phys., ITP--82--87P, Kiev 1982, 38~p. (in Russian).
\bibitem{Pryce}
Pryce M.~H.~L. The mass-centre in the restricted theory of relativity and its
connection with the quantum theory of elementary particles //
Proc.\ Roy.\  Soc.\  London,  1948, vol.~A195, No~1040, p.~62--81.
\bibitem{Flem}
Fleming G.~N. // Covariant position operators, spin, and locality
Phys.\ Rev., 1965, vol.~B137, p.~188--197.
\bibitem{Bak_Th}
Bakamjian B.\ and Thomas L.~H. Relativistic particle  dynamics.\ II
// Phys.\ Rev., 1953, vol.~92, No~5, p.~1300--1310.
\bibitem{Sk_1}
Sokolov S.~N.  Coordinates in relativistic Hamiltonian mechanics //
Teor.\ Mat.\ Fiz., 1985, vol.~62, No~2, p.~210--221 (in Russian).
\bibitem{Sk_2}
Sokolov S.~N. Relativistic addition of direct interactions in point form
of dynamics // Teor.\ Mat.\ Fiz., 1978, vol.~36, No~2, p.~193--207
(in Russian).
\bibitem{Dv_Kl}
Duviryak A.~A.\ and Kluchkovsky Yu.~B. Space-time interpretation of
relativistic Hamiltonian mechanics of particle system // Ukr.\ Fiz.\
Zh., 1992, vol.~37, No~2, p.~313--320 (in Ukrainian).
\bibitem{GTY_TM}
Gaida R.~P., Tretyak V.~I., and Yaremko Yu.~G. Center-of-mass variables
in the relativistic Lagrangian dynamics of a system of particles //
Teor.\ Mat.\ Fiz.\ 1994, vol.~101, No~ 3, p.~402--416 (in
Russian; transl.\ in English: Theor.\ Math.\ Phys., 1994, vol.~101,
No~3, p.~1433--1453).
\bibitem{Y_IPPMM}
Jaremko Yu.~G. Relativistic Lagrangian dynamics of $N$-particle system
in terms of centre-of-mass coordinates./
Preprint of the Institute of Applied Problems  of
Mechanics and Mathematics, Lviv 1988, No~14-88, 34~p.\ (in Russian).
\bibitem{Y_90}
Yaremko Yu.~G. Quasi-relativistic two-body problem in Lagrangian
centre-of-mass variables./ Preprint of the Institute for Theor.\
Phys., ITP--90--63U, Kiev 1990, 33~p.\ (in Ukrainian).
\bibitem{GTY_UPJ}
Gaida R.~P., Tretyak V.~I., and Yaremko Yu.~G. Center-of-mass variables
in the relativistic Lagrangian mechanics of particle system  //
Ukr.\ Fiz.\ Zh., 1991, vol.~36, No~12, p.~1807--1818 (in Ukrainian).
\bibitem{WR}
Weyssenhoff J.,\  Raabe A. Relativistic dynamics of spin-fluids and
spin-particles // Acta Phys.\ Pol., 1947, vol.~9, No~1, p.~7--18.
\bibitem{DS}
Damour T.\ and Schafer G. Redefinition of position variables
and the reduction of  higher-order  Lagrangians //
J.\ Math.\ Phys., 1991, vol.~32, No~1, p.~127-134.
\bibitem{JLM}
Llosa J.,\  Vives J.  Hamiltonian formalism for nonlocal Lagrangians //
J.\ Math.\ Phys., 1994, vol.~35, No~6, p.~2856--2877.
\bibitem{WH}
Woodcock H.~W.,\  Havas P. Approximately  relativistic  Lagrangians
for classical interacting point particles // Phys.\ Rev.\ D, 1972,
vol.~6, No~12, p.~3422--3444.
\bibitem{Bz}
Ba\.za\'nski S. Lagrange function for the motion of charged particles
in general relativity theory //
Acta  Phys.\  Pol., 1957, vol.~16, No~6, p.~423--433.
\bibitem{Kl}
Kluchkovsky Yu.~B.  Approximate Lorentz-invariance and
PPN-formalism in the theory of gravitation //
Math.\ Methods and Phys.-Mech.\ Fields, No~16, Kiev,
Nauk.\ Dumka, 1982, p.~107--110 (in Russian).
\bibitem{WM}
Whitrow G.~J., Morduch G.~E. Relativistic theories of gravitation.
-- In: Vistas in Astronomy, vol.~6. London,
Pergamon Press, 1965, p.~1--67.
\bibitem{Ms}
Messia A. Quantum Mechanics. Moscow, Nauka, 1979 (in Russian).
\bibitem{BLP}
Berestetskii V.~B., Lifshitz E.~M., and Pitaevskii L.~P. Quantum
Electrodynamics. Moscow, Nauka, 1989 (in Russian).

\end{thebibliography}
\end{document}